\documentclass[times,final]{elsarticle}

\usepackage{jcomp_af}
\usepackage{framed,multirow}

\usepackage{amssymb}
\usepackage{latexsym}
\usepackage{hyperref}
\usepackage{bigdelim}
\usepackage{tabularx}
\usepackage{booktabs}

\usepackage[singlelinecheck=false]{caption}

\usepackage{url}
\usepackage{xcolor}
\usepackage{algorithm}
\usepackage{algorithmicx}
\usepackage[fleqn]{amsmath}
\definecolor{newcolor}{rgb}{.8,.349,.1}

\DeclareMathAlphabet\mathbfcal{OMS}{cmsy}{b}{n}

\graphicspath{{../figs/}}

\journal{Manuscript for submission to the Journal of Computational Physics {(LLNL-JRNL-2003045)}}

\begin{document}

\verso{J. R. Angus \textit{et al}}

\begin{frontmatter}

\title{A Binary Collision Method for Screened Coulomb Collisions in weakly and moderately coupled Plasmas}

\author{Justin Ray Angus\corref{cor1}}
\cortext[cor1]{Corresponding author}
\ead{angus1@llnl.gov}
\author{Johannes Johgan van de Wetering}

\address{Lawrence Livermore National Laboratory, Livermore, CA 94550, USA}


\begin{abstract}
A binary-pairing Monte Carlo collision method is presented here for screened Coulomb collisions in plasmas that is valid in the weakly and moderately coupled regimes. The method models the Fokker-Planck collision operator with first order corrections with respect to the Coulomb logarithm in the appropriate limits and identically reduces to a standard Monte Carlo method for solving the Boltzmann collision integral for screened Rutherford collisions when the time step is sufficiently small. The method is exceptionally simple and can be integrated into existing binary methods for cumulative scattering with the addition of about 10 lines of code.
\end{abstract}


\end{frontmatter}


 
\section{Introduction}

The Landau-Fokker-Planck (LFP) equation governs collisional processes in fully ionized, weakly coupled plasmas, where the Coulomb logarithm is large ($\ln\Lambda\gtrsim 10$). This equation describes the collisional evolution of a species' distribution function due to cumulative small-angle scattering. The Coulomb logarithm quantifies the relative importance of cumulative small-angle scattering to single large-angle collisions. The latter can be important for moderately coupled plasmas where $2\lesssim \ln\Lambda\lesssim 10$, such as those found in inertial confinement fusion (ICF) experiments \cite{lindl2014}. Li and Petrasso generalized the LFP equation to include large-angle scattering \cite{Li1993_FPmod}, which they applied to quantify charged particle stopping powers in ICF plasmas \cite{Li1993}. In this model, contribution of large-angle scattering to the stopping power appears through terms that scale a $1{/}\ln\Lambda$. Recently, this model has been used to examine the effects of alpha-ion stopping on ignition criteria in ICF experiments \cite{reichelt2024}.

The binary-pairing algorithms by Takizuka and Abe \cite{Takizuka1977205} (TA77) and Nanbu \cite{nanbu1997} (N97) are two standard Monte Carlo (MC) methods used in particle-in-cell (PIC) codes to model the LFP equation. These methods belong to a broader class of MC methods that solve the LFP equation \cite{bobylev2000,Bobylev2013123}. Several studies have focused on incorporating large-angle collisions into these binary-pairing methods. Turrell et al. achieved this by splitting the scattering into small-angle and large-angle groups, separated by a cut off angle (typically 90 degrees) \cite{TURRELL2015144}. Large-angle scattering was modeled using the conventional MC method for the Boltzmann collision integral, while small-angle scattering was treated using the TA77 method with a reduced Coulomb logarithm. Higginson extended the N97 method to include large-angle collisions using a continuous distribution function that is like that from N97 for angles below a critical angle and transitions to the Rutherford distribution at larger angles \cite{HIGGINSON2017589}. This method requires solving a nonlinear three-variable system for each binary pair to enforce the imposed constraints on the distribution function for the scattering angle. 

This paper presents a binary-pairing MC method for screened Coulomb collisions in plasmas, applicable to both the weakly and moderately coupled regimes. The method models the LFP equation with first order $1/\ln\Lambda$ corrections in the appropriate limits and transitions to a standard MC method for solving the Boltzmann collision integral for screened Rutherford collisions when the time step is sufficiently small. The algorithm is remarkably simple and can be integrated into any existing MC method for cumulative scattering with the addition of about 10 lines of code. The key insight enabling this simplicity is that the specific form of the distribution function used to model the LFP equation is unimportant. What matters is that the transport moment matches that obtained from the Rutherford differential cross section for screened Coulomb collisions \cite{Bobylev2013123}.

The remainder of the paper is outlined as follows. Methods for modeling screened Coulomb collisions in a plasma via cumulative and single scattering events are discussed in Sec.\,\ref{sec:Theory}. Our generalized method for screened Coulomb collisions in plasmas is presented in Sec.\,\ref{sec:GMC}. Verification tests are presented in Sec.\,\ref{sec:verification}. In Sec.\,\ref{sec:alphaBeam}, the method is used to quantify knock-on ion production and collisional stopping of a high-energy alpha beam in a DT plasma at ignition-relevant conditions \cite{zylstra2022}. Finally, further discussion and a summary is provided in Sec.\,\ref{sec:summary}.

\section{The Rutherford cross section for screened Coulomb collisions in plasmas} \label{sec:Theory}

Consider a particle $i$ belonging to species $\alpha$ with charge $q_{\alpha}=Z_{\alpha}e$ and mass $m_{\alpha}$ and particle $j$ belonging to species $\beta$ with charge $q_{\beta}=Z_{\beta}e$ and mass $m_{\beta}$. The differential cross section for a Coulomb collision between particle $i$ and $j$ is given by the Rutherford formula:
\begin{equation}
I_{R} = \frac{b^2_{\bot}}{4\sin^4\frac{\theta}{2}}, \label{Eq:Rutherford}
\end{equation}
where $\theta$ is the polar scattering angle in the center-of-momentum frame, and $b_{\bot} = |q_{\alpha}q_{\beta}|/\left(4\pi\epsilon_0\mu_{\alpha\beta}v_{ij}^2\right)$ is the classical impact parameter for a $90^{\circ}$ scatter with $\mu_{\alpha\beta}=m_{\alpha}m_{\beta}/\left(m_{\alpha}+m_{\beta}\right)$ the reduced mass and $v_{ij}$ is the relative velocity. The total Rutherford cross section is infinite. However, inter-particle potentials in a plasma are screened at a distance set by the plasma Debye length, $\lambda_{De}$. This gives a maximum impact parameter, $b_{\max}=\lambda_{De}$, which corresponds to a minimum scattering angle. The total Rutherford cross section for Coulomb collisions in a plasma is
\begin{equation}
\sigma_{\text{tot}} \equiv 2\pi\int_{\theta_{\min}}^{\pi}I_R\sin\theta d\theta = \pi b_{\bot}^2\left(\frac{1}{\sin^2\frac{\theta_{\min}}{2}}-1\right). \label{Eq:sigma_tot}
\end{equation}
The corresponding transport cross section is
\begin{equation}
\sigma_{\text{tr}} \equiv 2\pi\int_{\theta_{\min}}^{\pi}\left(1-\cos\theta\right)I_R\sin\theta d\theta = 4\pi b_{\bot}^2\ln\Lambda, \label{Eq:sigma_tr}
\end{equation}
where 
\begin{equation}
\ln\Lambda = \frac{1}{2}\ln\left[\frac{1}{\sin^2\frac{\theta_{\min}}{2}}\right] \label{Eq:Clog}
\end{equation}
is the Coulomb logarithm.

The $\sin^2\frac{\theta_{\min}}{2}$ terms in Eqs.\,\ref{Eq:sigma_tot}-\ref{Eq:Clog} can be expressed in terms of the impact parameter $b$. Classically, the relationship between the scattering angle $\theta$ and the impact parameter $b$ is given as
\begin{equation}
\sin^2\frac{\theta}{2} = \frac{b^2_{\bot}}{b^2_{\bot} + b^2}. \label{Eq:sinsq_classical}
\end{equation}
A fundamental impact parameter from quantum mechanics is $b_{\text{qm}} = \hbar/\left(2\mu_{\alpha\beta}v_{ij}\right)$. The classical relation given in Eq.\,\ref{Eq:sinsq_classical} is only valid for low-velocity interactions such that $\kappa \equiv b_{\bot}/b_{\text{qm}}>1$. In this work, we use the following generalization of Eq.\,\ref{Eq:sinsq_classical} obtained using the transformations $b^2_{\bot}\rightarrow b^2_{\bot}+b^2_{\text{qm}}$ and $b^2\rightarrow \left(b+b_{\text{qm}}\right)^2-b^2_{\text{qm}}$ \cite{lindhard1996}:
\begin{equation}
\sin^2\frac{\theta}{2} = \frac{b^2_{\bot} + b^2_{\text{qm}}}{b^2_{\bot} + \left(b+b_{\text{qm}}\right)^2}, \label{Eq:sinsq_qm}
\end{equation}
which results in the following expression for the Coulomb logarithm:
\begin{equation}
\ln\Lambda = \frac{1}{2}\ln\left[\frac{b^2_{\bot} + \left(b_{\max}+b_{\text{qm}}\right)^2}{b^2_{\bot} + b^2_{\text{qm}}}\right]. \label{Eq:Clog2}
\end{equation}
This expression reduces to the classical value for large $\kappa$ \cite{TURRELL2015144} and agrees well with the quantum calculation of the Coulomb logarithm otherwise \cite{lindhard1996}.

The normalized mean free path associated with the total and transport cross sections are fundamental parameters for scattering algorithms. For particle $i\in\alpha$ colliding with particle $j\in\beta$ during interval $\Delta t$, these parameter are defined, respectively, as 
\begin{eqnarray}
N_{ij} &=& \sigma_{\text{tot}}v_{ij}n_{\min}\Delta t, \label{Eq:Nij}\\
s_{ij} &=& \sigma_{\text{tr}}v_{ij}n_{\min}\Delta t, \label{Eq:sij}
\end{eqnarray}
where $n_{\min}=w\min\left(N_{\alpha},N_{\beta}\right)/\Delta V$ is the density of the species with the lower number of simulation particles in a cell of volume $\Delta V$. The simulation particles are assumed to have equal weights in this work ($w_i=w_j=w$), in which case momentum and energy are identically conserved for each binary pair, and the expressions in Eq.\,\ref{Eq:Nij}-\ref{Eq:sij} are symmetric with respect to $i$ and $j$. The particle pairing procedure used for the algorithm here is the same as that used in TA77 - the number of binary pairs per time step is $\max\left(N_{\alpha},N_{\beta}\right)$ and the velocity of particles that collide multiple times are updated sequentially. See Ref.\cite{angus2024mpmccc} for a discussion on how to properly account for particles of varying weights.

In the standard MC method for solving the Boltzmann collision integral, the probability of a collision occurring between a binary pair of particles in time step $\Delta t$ is given by $1-\exp\left(-N_{ij}\right)$. If a collision occurs, the polar scattering angle is determined from Eq.\,\ref{Eq:sinsq_qm} with $\tilde{b}^2\equiv\left(b+b_{\text{qm}}\right)^2$ obtained from setting the cumulative distribution function to $\mathcal{R}\in\left[0,1\right]$:
\begin{equation}
\mathcal{R} = \frac{2\pi}{\sigma_{\text{tot}}}\int_{\theta_{\min}}^{\theta}I_R\sin\theta d\theta = \frac{\tilde{b}_{\max}^2-\tilde{b}^2}{\tilde{b}^2_{\max}-b^2_{\text{qm}}}. \label{eq:cumdist}
\end{equation}

\subsection{Angular distributions for cumulative and single scattering} \label{sec:LFP_models}

$N_{ij}$  in Eq.\,\ref{Eq:Nij} represents the expected number of Rutherford collisions occurring in $\Delta t$. The normalized transport mean free path can also be expressed as $s_{ij} = \left<1-\cos\theta\right>_1N_{ij}$, where $\left<\cdot\right>_1$ denotes the expected value for a single collision. For cumulative small angle collisions, $\left<1-\cos\theta\right>_1$ is small in which case $\left<1-\cos\theta\right>_1 \approx \frac{1}{2}\left<\theta^2\right>_1$. Given that the events are random and independent, and that the azimuthal angle $\phi$ is uniformly distributed, this is a 2D diffusive process. By the central limit theorem, for large $N_{ij}$, the polar scattering angle $\theta$ will follow the Rayleigh distribution with a variance of $2s_{ij}$, which conforms to the probability distribution function described in N97.

Cumulative scattering models are used for situations where it is not feasible to use a time step that resolves individual collisions; $N_{ij}\gg1$. Each particle collides with at least one particle from each species each time step and the polar scattering angle is sampled from some probability distribution function. While the probability distribution function given in N97 is physically expected for cumulative Coulomb scattering in plasmas, it is not necessary to sample from this distribution to correctly reproduce Fokker-Planck transport in a plasma. In order for a binary-pairing method to be a numerical solution to the LFP equation, all that is required of the probability distribution function for the polar scattering angle, $P_{ij}\left(\theta\right)$, is that it satisfies the following conditions \cite{Bobylev2013123}:
\begin{eqnarray}
\lim_{s_{ij}\rightarrow 0}\int_0^{\pi}\left(1-\cos\theta\right)^kP_{ij}\left(\theta\right)d\theta &=& s_{ij}, \ \text{for } \ k=1, \label{Eq:Pcond_k1} \\
&=& 0, \ \ \ \text{for } \ k\geq 2. \label{Eq:Pcond_k2}
\end{eqnarray}
That is, the expected value of $\left(1-\cos\theta\right)$ must equal the normalized transport mean free path, and higher transport moments must tend to zero faster as $s_{ij}\rightarrow 0$. The probability distribution function for the scattering angle from TA77, N97, and that in Ref.\,\cite{Bobylev2013123} (B13) are given, respectively, as
\begin{eqnarray}
\text{TA77:}&& \ \ P_{ij}\left(\theta\right) = \frac{1}{\sqrt{\pi s_{ij}}}\frac{1}{\cos^2\frac{\theta}{2}}\exp\left(-\frac{\tan^2\frac{\theta}{2}}{s_{ij}}\right), \label{Eq:P_TA77} \\
\text{N97:}&& \ \ P_{ij}\left(\theta\right)  = \frac{A_{ij}\sin\theta}{2\sinh\left(A_{ij}\right)}\exp\left(A_{ij}\cos\theta\right), \ \text{where } \ \text{coth}\left(A_{ij}\right) - \frac{1}{A_{ij}} = \exp\left(-s_{ij}\right), \label{Eq:P_N97} \\
\text{B13:}&& \ \ P_{ij}\left(\theta\right)  = \delta\left[1-\cos\theta-\min\left(s_{ij},2\right)\right]\sin\theta. \label{Eq:P_B13}
\end{eqnarray}
Each of these methods satisfy the conditions given in Eq.\,\ref{Eq:Pcond_k1}-\ref{Eq:Pcond_k2}, and are thus valid distribution functions to use to model the LFP collision integral in the small time step limit. The formula for the scattering angle can be obtained by setting the cumulative distribution of $P_{ij}\left(\theta\right)$ equal to a uniform random number $\mathcal{R}\in\left[0,1\right]$. The scattering angle for the B13 method is $\cos\theta = 1-\min\left(s_{ij},2\right)$, which doesn't require a random number. 

\begin{figure}[!ht] 
\centering
\includegraphics[scale=1.0]{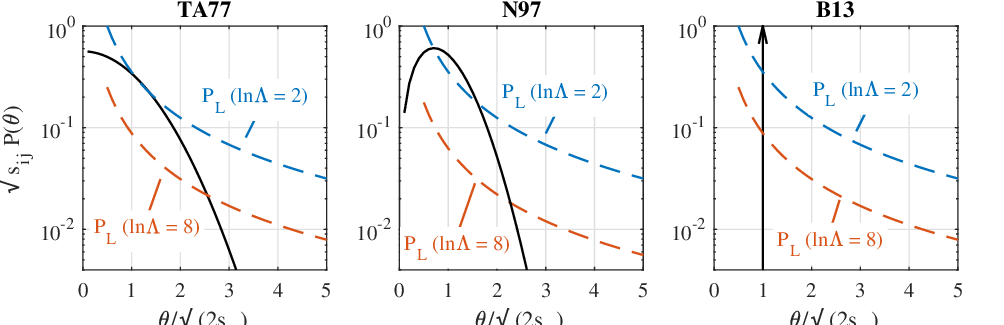}
\caption{Angular distributions in the small angle limit for TA77 (left panel), N97 (middle panel), and B13 (right panel). The probability distribution function for single Rutherford collisions is shown for $\ln\Lambda = 2$ and $\ln\Lambda = 8$ with the dashed blue and red curves, respectively.} \label{fig:angularDistributions}
\end{figure}

The probability distribution for a single Rutherford collision is by definition $I_R2\pi\sin\theta/\sigma_{\text{tot}}$. The total probability distribution for a single Rutherford collision in $\Delta t$ is this value multiplied by $N_{ij}$:
\begin{equation}
P_{ij,R}\left(\theta\right) = N_{ij}\frac{I_R2\pi\sin\theta}{\sigma_{\text{tot}}} = \frac{s_{ij}}{4\ln\Lambda}\frac{\cos\frac{\theta}{2}}{\sin^3\frac{\theta}{2}}. \label{Eq:PR}
\end{equation}
The probability distribution functions for the three cumulative models given in Eqs.\,\ref{Eq:P_TA77}-\ref{Eq:P_B13} are shown in Fig.\,\ref{fig:angularDistributions} and compared with that for single Rutherford scattering using $\ln\Lambda = 2$ and $\ln\Lambda = 8$. The small angle approximation is used for these figures. When using a cumulative probability distribution, the probability of scattering into angles larger than a few times the mean square scattering angle is greatly underestimated (see Sec.\,13.6 of Ref.\,\cite{jackson1998classical}). The relative importance of these larger angle scattering events increases as $\ln\Lambda$ decreases.

\section{Generalized method for screened Coulomb collisions in plasmas} \label{sec:GMC}

In Ref.\,\cite{HIGGINSON2017589}, the N97 method was extended to include large-angle collisions using a continuous distribution for the scattering angle that is like that from N97 for angles below a critical angle and transitions to the Rutherford distribution at larger angles. The constraints for the distribution are 1) the total probability is one, 2) and total variance matches the expression given in N97, and 3) the distribution function is continuous with $\theta$. This requires solving a nonlinear three variable system for each binary pair to find the amplitude and variance of the cumulative distribution and the critical angle where the cumulative and single scattering distributions intersect. We present an alternative method to combine cumulative and single scattering distributions that is independent of which cumulative distribution is used and doesn't require solving a nonlinear system. The key insight enabling the development of this model is that the shape of the distribution function used for the cumulative part does not matter. All that matters for reproducing Fokker-Planck transport in the appropriate limits is that the distribution satisfies the conditions given in Eq.\,\ref{Eq:Pcond_k1}-\ref{Eq:Pcond_k2}. Thus, we do not impose the constraint that the probability distribution function must be continuous with scattering angle.

The distribution we consider in this work can formally be expressed as:
\begin{equation}
P_{ij}\left(\theta\right) = \alpha P_{ij,M}\left(s_{ij,M},\theta\right) + \frac{s_{ij}}{4\ln\Lambda}\frac{\cos\frac{\theta}{2}}{\sin^3\frac{\theta}{2}}H\left(\theta-\theta_c\right), \label{Eq:Ptot}
\end{equation}
where $H\left(\theta-\theta_c\right)$ is the Heaviside step function and $\theta_c$ is the cut off angle for single Rutherford collisions. The subscript $M$ is used to denote multiple (i.e., cumulative) scattering. The parameters $\alpha$ and $s_{ij,M}$ for the cumulative part of the distribution in Eq.\,\ref{Eq:Ptot} are obtained from the constraints that the total probability is unity and the transport moment is $s_{ij}$ for $s_{ij}\rightarrow 0$ (see Eq.\,\ref{Eq:Pcond_k1}).

The total probability of a single Rutherford collision between $\theta_c$ and $\pi$ is
\begin{equation}
S_R = \int^{\pi}_{\theta_c} \frac{s_{ij}}{4\ln\Lambda}\frac{\cos\frac{\theta}{2}}{\sin^3\frac{\theta}{2}}d\theta = N_{ij}\frac{\tilde{b}^2_c - b_{\text{qm}}^2}{\tilde{b}_{\max}^2 - b_{\text{qm}}^2}, \label{Eq:SR}
\end{equation}
where $\tilde{b}_c=b_c+b_{\text{qm}}$ is the effective impact parameter associated with $\theta_c$ via Eq.\,\ref{Eq:sinsq_qm}. Since the total probability is one, we have $\alpha = 1 - S_R$. The normalized transport mean free path for the cumulative part of the distribution, $s_{ij,M}$, is obtained from the requirement that the model is a numerical solution to the LFP equation in the appropriate limit. That is, the expected value of $1-\cos\theta$ using Eq.\,\ref{Eq:Ptot} must be $s_{ij}$ in the limit where $s_{ij}\rightarrow 0$:
\begin{equation}
\int_{0}^{\pi}\left(1-\cos\theta\right)P_{ij}\left(\theta\right)d\theta = \alpha s_{ij,M} + \frac{s_{ij}}{\ln\Lambda}\frac{1}{2}\ln\left[\frac{b^2_{\bot}+\tilde{b}_c^2}{b^2_{\bot}+ b^2_{\text{qm}}}\right] = s_{ij} \Rightarrow s_{ij,M} = \frac{s_{ij}}{1-S_R}\frac{\ln\Lambda_M}{\ln\Lambda}, \label{Eq:S12M}
\end{equation}
where we have assumed small $s_{ij,M}$ for the cumulative part in Eq.\,\ref{Eq:Ptot} and $\ln\Lambda_M\equiv\frac{1}{2}\ln\left[\left(b^2_{\bot}+\tilde{b}^2_{\max}\right)/\left(b^2_{\bot}+\tilde{b}^2_{c}\right)\right]$. To close the system, we need to specify how to determine $\tilde{b}_c$ for single Rutherford collisions. We choose to hold the probability of a single Rutherford collision fixed at $S_R = 0.1$ for $N_{ij}>0.1$. That is, we set
\begin{equation}
S_R = \min\left(N_{ij},0.1\right). \label{Eq:SR2}
\end{equation}
The cut off impact parameter $\tilde{b}_c$ is then determined directly from Eq.\,\ref{Eq:SR}.

The algorithm described above is summarized in Algorithm\,\ref{alg:GeneralCoulomb}, which describes how to determine the polar scattering angle in the center-of-momentum frame for a given binary pair of particles. All other aspects of the scattering method, such as how to form the binary pairs and how to update their velocities once the polar scattering angle is determined, are not discussed here as they are identical to that described elsewhere \cite{Takizuka1977205,angus2024mpmccc}. Step 12 in Algorithm\,\ref{alg:GeneralCoulomb} assumes that the B13 model is used for the cumulative scattering. However, the method works with any of the commonly used cumulative distributions functions for cumulative scattering given in Eqs.\,\ref{Eq:P_TA77}-\ref{Eq:P_B13}.

\begin{algorithm}
\caption{Algorithm to compute the polar scattering angle for a binary pair of particles $i$ and $j$.}
\begin{algorithmic}[1]
\State\hspace{\algorithmicindent} Compute the normalized transport mean free path using Eq.\,\ref{Eq:sij}: $s_{ij}$.
\State\hspace{\algorithmicindent} Compute the expected number of total Rutherford collisions: $N_{ij}=s_{ij}\sigma_{\text{tot}}/\sigma_{\text{tr}}$.
\State\hspace{\algorithmicindent} Set the probability for a single Rutherford scatter using Eq.\,\ref{Eq:SR2}: $S_R = \min(N_{ij},0.1)$.
\State\hspace{\algorithmicindent} Compute the cut off impact parameter squared using Eq.\,\ref{Eq:SR}: $\tilde{b}_c^2 = b^2_{\text{qm}} + \left(\tilde{b}^2_{\max}-b_{\text{qm}}^2\right)S_R/N_{ij}$.
\State\hspace{\algorithmicindent} Get a random number: $\mathcal{R}\in\left[0,1\right]$.
\State\hspace{\algorithmicindent} If $\mathcal{R}<=S_R$ (single Rutherford scatter) 
\State\hspace{\algorithmicindent}\hspace{\algorithmicindent} $\tilde{b}^2 = \tilde{b}_c^2 - \mathcal{R}/S_R\left(\tilde{b}_c^2 - b^2_{\text{qm}}\right)$.
\State\hspace{\algorithmicindent}\hspace{\algorithmicindent} $\cos\theta = \left(\tilde{b}^2-2b^2_{\text{qm}}-b^2_{\bot}\right)/\left(\tilde{b}^2+b^2_{\bot}\right)$.
\State\hspace{\algorithmicindent} Else if $N_{ij}>0.1$ (cumulative scatter) 
\State\hspace{\algorithmicindent}\hspace{\algorithmicindent} $\ln\Lambda_M = \frac{1}{2}\log\left[\left(b_{\bot}^2+\tilde{b}^2_{\max}\right)/\left(b_{\bot}^2+\tilde{b}^2_{c}\right)\right]$.
\State\hspace{\algorithmicindent}\hspace{\algorithmicindent} $s_{ij,M} = s_{ij}\ln\Lambda_M/\ln\Lambda/0.9$.
\State\hspace{\algorithmicindent}\hspace{\algorithmicindent} $\cos\theta = 1-\min\left(s_{ij,M},2\right)$.
\State\hspace{\algorithmicindent} Else $\cos\theta = 1$ (no scatter).
\end{algorithmic}  \label{alg:GeneralCoulomb}
\end{algorithm}

\begin{figure}[!ht] 
\centering
\includegraphics[scale=1.0]{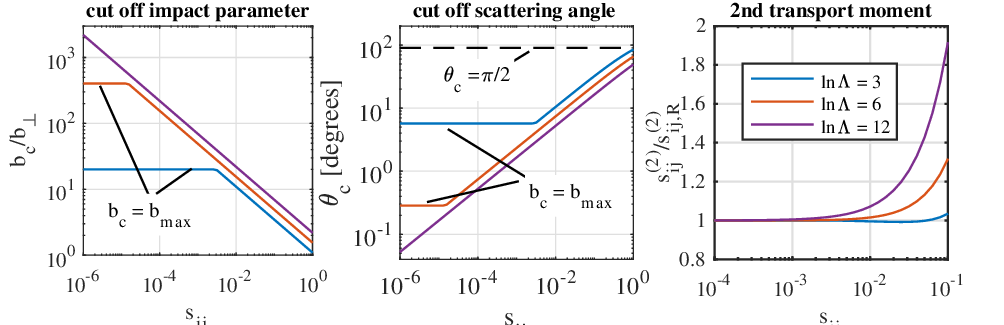}
\caption{Cut off impact parameter (left) and corresponding cut off scattering angle (middle) for a single Rutherford scatter vs the normalized transport mean free path using $S_R = \min(N_{ij},0.1)$ and $b_{\text{qm}}=0$. The B13 model is used for the results in these figures. Results are shown for $\ln\Lambda=2$ (blue), $\ln\Lambda=6$ (red), and $\ln\Lambda=12$ (purple). The right panel shows the second transport moment normalized to that for pure Rutherford scattering. The legend on the right plot applies to all three plots.} \label{fig:cutoffDiagram}
\end{figure}

The cut off impact parameter for this method versus $s_{ij}$ is shown in the left panel of Fig.\,\ref{fig:cutoffDiagram} for various values of $\ln\Lambda$ and using $b_{\text{qm}}=0$. The corresponding cut off angle for single Rutherford collisions is shown in the middle panel of this figure.  Values of $s_{ij}$ less than 0.01 are required to reproduce collisional transport processes with high accuracy. From Fig.\,\ref{fig:cutoffDiagram} it is seen that the $\theta_c<10^{\circ}$ for $s_{ij}=0.01$ and $\ln\Lambda\geq 3$, which is much smaller than $\theta_c=90^{\circ}$ that is used in the Ref.\,\cite{TURRELL2015144}. One way to look at this model is that we are doing single Rutherford scatterings in the largest range of impact parameters that is reasonably possible for a given simulation time step. The cumulative part is just there to ensure that the total transport mean free path is preserved when $N_{ij}>0.1$. Note that by definition $s_{ij,M}\rightarrow 0$ as $b_c\rightarrow b_{\max}$, which occurs as $N_{ij}\rightarrow 0.1$. So, for $N_{ij}<0.1$, the method here reduces identically to a conventional discrete Monte Carlo method for modeling the Boltzmann collision integral for screened Rutherford collisions.

Leading-order corrections to Fokker-Planck transport enter through terms that scale with the second transport moment \cite{Bobylev2013123}. Using the B13 model for the cumulative part of Eq.\,\ref{Eq:Ptot}, the second normalized transport mean free path associated with $P_{ij}$ in Eq.\,\ref{Eq:Ptot} is
\begin{equation}
s^{\left(2\right)}_{ij} \equiv \int_0^{\pi}\left(1-\cos\theta\right)^2P_{ij}d\theta =  \frac{s^2_{ij}}{\alpha}\left(\frac{\ln\Lambda_M}{\ln\Lambda}\right)^2 + s_{ij}\frac{\cos\theta_c+1}{2\ln\Lambda}. \label{Eq:sij2}
\end{equation}
The first term is zero for $s_{ij}$ such that $b_c=b_{\max}$, in which case $\theta_c=\theta_{\min}$ and $s_{ij}^{\left(2\right)}$ equals that obtained for screened Rutherford collisions. For $b_c<b_{\max}$, the first term of $s_{ij}^{\left(2\right)}/s_{ij}$ tends to zero like $s_{ij}$ as $s_{ij}\rightarrow 0$, and the second term saturates to a constant equal to $\frac{1}{2}\left(\cos\theta_c+1\right)/\ln\Lambda$. This method is thus a valid numerical method for solving the LFP equation with first order $1/\ln\Lambda$ corrections in the limit where $s_{ij}\rightarrow 0$.

For sufficiently small time steps and for $\theta_{\min}\ll 1$, relative corrections to Fokker-Planck transport in our model scale with $s^{\left(2\right)}_{ij,R}/s_{ij} = 1/\ln\Lambda$, in agreement with the Li and Petrasso model in Ref.\,\cite{Li1993_FPmod}. The second transport moment normalized to that for pure Rutherford scattering, $s_{ij,R}^{\left(2\right)}=\frac{1}{2}s_{ij}\left(\cos\theta_{\min}+1\right)/\ln\Lambda$, is shown in the right panel of Fig.\,\ref{fig:cutoffDiagram} vs. $s_{ij}$ for various values of $\ln\Lambda$. For moderately coupled plasmas, $s_{ij}\lesssim 0.01$ is more than sufficient to accurately capture these corrections.

\section{Method Verification tests} \label{sec:verification}

The scattering method described in Algorithm\,\ref{alg:GeneralCoulomb}, which we refer to as the generalized Coulomb method (GCM), is implemented in the PICNIC code \cite{angus2022,angus2023,angus2024_axisymmetric,angus2024mpmccc}. Results of various verification tests are presented here to illustrate the correct implementation of the model in PICNIC and the physical accuracy of the method. The Debye length used for $b_{\max}$ in PICNIC is defined as $\lambda_{De}^{-2}=\sum_{s}\lambda_{De,s}^{-2}$, where $\lambda_{De,s}$ is the Debye length for species $s$. Furthermore, while the analysis presented previously is non-relativistic, the simulation results are obtained using a fully relativistic implementation \cite{perez2012}. Relativistic generalizations of relevant quantities are given in Appendix A.

The tests are performed using a 2D numerical grid with $20{\times}20$ cells, but particle motion and forces are turned off making the simulations 0D (velocity space only). The results shown are averaged over the grid cells, which is the same as performing $400$ 0D simulations with the specified number of particles per cell and taking the average. For each test, the initial particle velocities are obtained by sampling from a shifted Maxwellian distribution determined by the population's initial temperature and mean velocity.

\subsection{Test 1: ion-ion thermalization}

\begin{figure}[!ht] 
\centering
\includegraphics[scale=1.0]{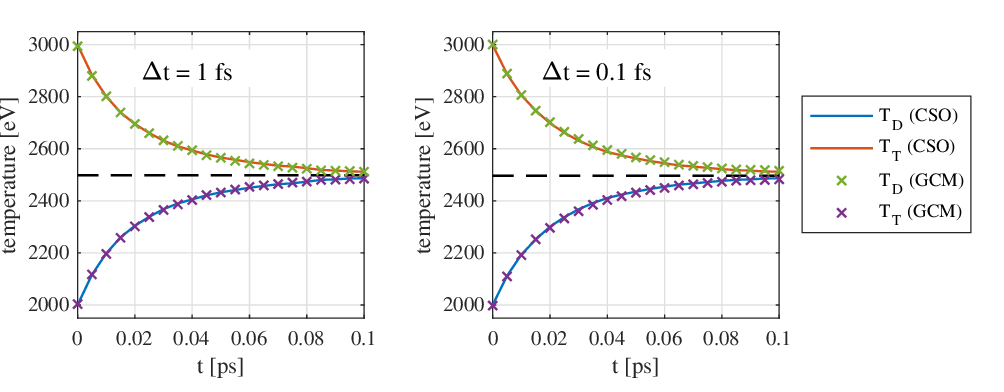}
\caption{Ion temperature relaxation tests. $n_D=n_T=2.5{\times}10^{31}$/m$^3$. $N_{ppc} = 100$ for each species. Solid curves are obtained using cumulative scattering only (CSO). Results obtained using the GCM (Algorithm\,\ref{alg:GeneralCoulomb}) are shown with the x-markers. The simulation time step is $\Delta t = 1$\,fs and $\Delta t = 0.1$\,fs, for the left and right panels, respectively.} \label{fig:ionRelaxation}
\end{figure}

Thermal relaxation of a 50/50 DT plasma with $n_D=n_T=2.5{\times}10^{31}$/m$^3$ is considered for the first test. The initial temperature is $T_D=3$\,keV for the deuterium ions and $T_T=2$\,keV for the tritium ions. Time profiles of the species temperatures from simulations using cumulative scattering only (CSO) are compared with those using the GCM in Fig.\,\ref{fig:ionRelaxation}. Results are shown using simulation time steps of $\Delta t = 1$\,fs and $\Delta t = 0.1$\,fs. The results using the GCM are indistinguishable from those obtained using CSO. The B13 model is used for the CSO simulations and for the cumulative part of the GCM. Using the initial conditions, the Coulomb logarithm for D-T collisions is $\ln\Lambda\approx 5$ and the characteristic thermalization time is about $30$\,fs \cite{richardson2019}. The mean value of $s_{ij}$ for D-T collisions from these simulations is about 0.2 for $\Delta t = 1$\,fs and 0.02 for $\Delta t = 0.1$\,fs. From Fig.\,\ref{fig:cutoffDiagram}, it can be inferred that the mean value of the cut off angle for single Rutherford scattering is about 10 degrees for $\Delta t=1$\,fs and about 2 degrees for $\Delta t=0.1$\,fs.

\subsection{Test 2: knock-on ion energy distribution}

For our second test, we directly validate the knock-on ion production by a mono-energetic alpha beam colliding with D and T ions. The number of knock-on ions produced (per unit volume per unit time per unit knock-on ion energy) via Coulomb scattering for a mono-energetic alpha beam with density $n_{\alpha}$ and energy $E_{\alpha}=\frac{1}{2}m_{\alpha}v^2_{\alpha}$ incident on a cold ion species ($T_i=0$) with density $n_i$ (see Eq.\,11 of Ref.\,\cite{TURRELL2015144}) is
\begin{equation}
Q_i\left(E_i'\right) = \left(\frac{q_iq_{\alpha}}{4\pi\epsilon_0}\right)^2\frac{2\pi n_in_{\alpha}}{v_{\alpha}m_{i}E_i'^2}, \label{Eq:Qi}
\end{equation}
where
\begin{equation}
E_i' = \frac{2\mu^2_{i\alpha}}{m_im_{\alpha}}E_{\alpha}\left(1-\cos\theta\right). \label{Eq:Ei}
\end{equation}
is the post-scatter energy of the target ion species.

\begin{figure}[!ht] 
\centering
\includegraphics[scale=1.0]{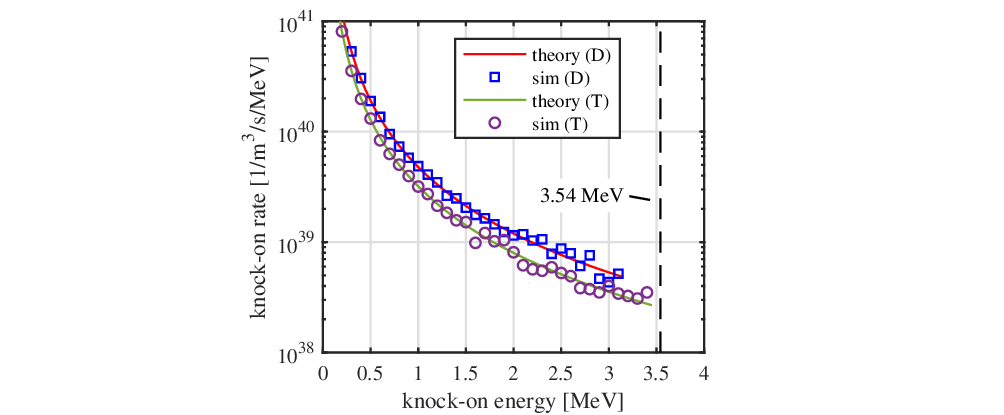}
\caption{Verification of implementation of large-angle scattering model for Coulomb collisions in PICNIC. These figures show the energy distribution of the rate of knock-on ions generated from a mono-energetic alpha beam with $E_{\alpha}=3.54$\,MeV interacting with a cold D or T plasma. The simulation time step is $\Delta t = 0.1$\,fs.} \label{fig:KnockOnVerify_Coulomb}
\end{figure}

For the simulations we consider an alpha beam with energy $E_{\alpha} = 3.54$\,MeV and density $n_{\alpha}=5{\times}10^{29}$/m$^3$. The ions have $T_i=0$ and $n_i=5{\times}10^{31}$/m$^3$. One simulation is performed using $i=D$ and another for $i=T$. The energy distribution of the knock-on ion generation rate is determined from the simulations by running many simulations for a single time step with only collisions between the alpha beam and the target ion species included. The simulation results are shown in Fig.\,\ref{fig:KnockOnVerify_Coulomb} to be in agreement with the theoretical expectation from Eq.\,\ref{Eq:Qi}. The knock-on generation rate for T ions is slightly smaller than for D ions, but the maximum energy corresponding to $\theta=\pi$ is slightly larger, as expected from Eqs.\,\ref{Eq:Qi}-\ref{Eq:Ei}.

\section{Slowing down of a high-energy alpha beam in a DT plasma under ignition relevant conditions} \label{sec:alphaBeam}

In this section we consider the collisional stopping of a high-energy alpha beam in a 50/50 DT plasma using ignition relevant conditions \cite{Li1993,TURRELL2015144,iwata2024_APSDPP}. The mono-energetic alpha beam is initialized with $E_{\alpha}=3.54$\,MeV and $n_{\alpha}=5{\times}10^{29}$/m$^3$. The D and T densities are $n_{D}=n_T=5{\times}10^{31}$/m$^3$. The electron density is $n_e=5.1{\times}10^{31}$/m$^3$. The initial temperature of the electrons and ions is $T_e=T_D=T_T=3$\,keV. The default number of particles per cell for each species used in the simulations is $20/1000/1000/2040$ for $\alpha$, D, T, and electron species, respectively. The default simulation time step is $\Delta t = 0.1$\,fs. 

\begin{figure}[!ht] 
\centering
\includegraphics[scale=1.0]{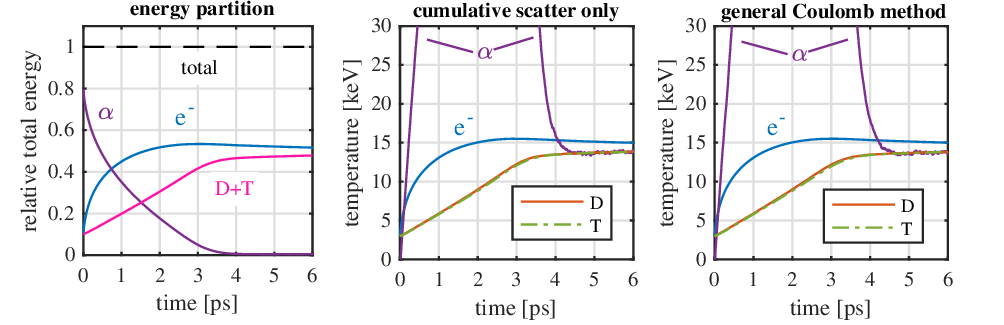}
\caption{The partition of total energy between the species vs. time is shown in the left panel. These results are obtained using the GCM and are indistinguishable from results obtained using CSO (not shown). Species temperatures vs. time are shown in the middle (using CSO) and right (using the GCM) panels. The simulations use the default time step of $\Delta t =0.1$\,fs and number of particle per cell.} \label{fig:alphaBeam_energyEvolution}
\end{figure}

The alpha beam slows down and thermalizes as it collisionally transfers energy to the electrons and ions in the plasma. The time evolution of the energy partition between the different species and their temperatures for this process are shown in Fig.\,\ref{fig:alphaBeam_energyEvolution}. Initially, the high-energy alpha particle stop mostly on the electrons, causing them to rapidly heat from $3$\,keV to about $10$\,keV after about $400$\,fs. The electrons reach a peak temperature of about 15\,keV after 3\,ps. The alpha particles have completely thermalized with the D/T ions just after 4\,ps with a common temperature of just under 14\,keV.

Results obtained using a cumulative scattering only model (i.e., Fokker-Planck) are also shown in Fig.\,\ref{fig:alphaBeam_energyEvolution}. The results are indistinguishable from those obtained using the general Coulomb method that includes large-angle scattering. This result can be understood using the stopping power formula in Ref.\,\cite{Li1993}. The large-angle scattering corrections to this formula scale as $m_f/\left(m_{\alpha}\ln\Lambda\right)$, where $m_f$ is the mass of the field species. Using the initial conditions, the characteristic values for $\ln\Lambda$ for alpha-electron interactions is about three and that for alpha-D/T is about 10, consistent with values in Fig.\,1 of Ref.\,\cite{Li1993}. The alpha particles mostly stop on the electrons, but because of the scaling with $m_f/m_{\alpha}$, we do expect noticeable differences for alpha-electron stopping when including large-angle scattering. Furthermore, because $m_f/\left(m_{\alpha}\ln\Lambda\right)\approx 1/20$ for alpha-D/T, corrections to the alpha-ion stopping only about a $5\%$ effect. In other words, for these conditions, alpha stopping is predominately governed by the transport cross section, which is the same independent of whether large-angle scattering is included.

\begin{figure}[!ht] 
\centering
\includegraphics[scale=1.0]{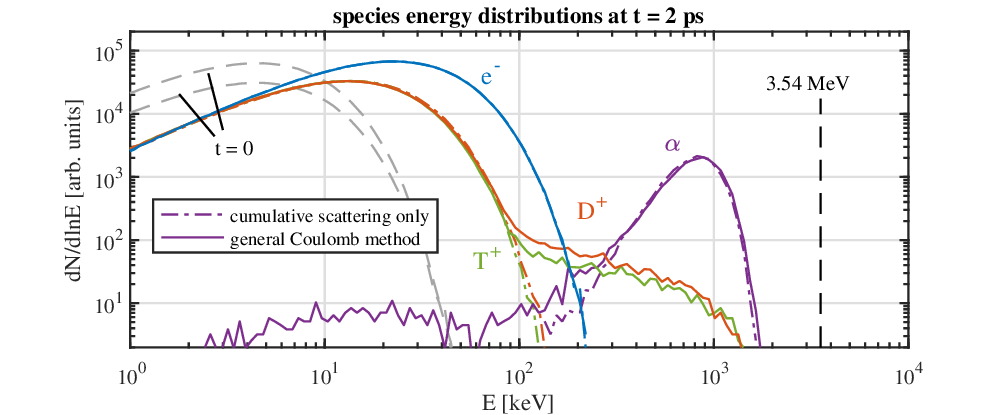}
\caption{Species energy distributions, $dN/d{\ln\left(E\right)} = E^{3/2}f(E)$, at $t = 2$\,ps from simulations using cumulative scattering only (dashed-dot lines) and using the general Coulomb method (solid lines) The grey lines illustrate the initial energy distributions at $t = 0$ for the electrons and D/T.} \label{fig:alphaBeam_EDF_2ps}
\end{figure}

The main effect of including large-angle scattering is the generation of high-energy knock-on D/T ions. The energy distribution of each species at $t=2$\,ps is shown in Fig.\,\ref{fig:alphaBeam_EDF_2ps}. The dashed-dotted curves are obtained using cumulative scattering only. The solid lines are obtained using the generalized Coulomb method via Algorithm\,\ref{alg:GeneralCoulomb}. The results obtained using the general Coulomb method display a large enhancement of the high energy tail of the distribution for the D/T ions in the range of $10^2-10^3$\,keV. Although high-energy knock-on ions make up only a small fraction of the overall population, they can enhance fusion reaction rates and affect the energy spectrum of D-T fusion neutrons in burning plasmas \cite{jeet2024}. The trends for the high-energy D/T ions in Fig.\,\ref{fig:alphaBeam_EDF_2ps} align with those from the knock-on verification tests in Fig.\,\ref{fig:KnockOnVerify_Coulomb}: the number of knock-on T ions is slightly lower than that of D ions, but their maximum energy is slightly larger.

\begin{figure}[!ht] 
\centering
\includegraphics[scale=1.0]{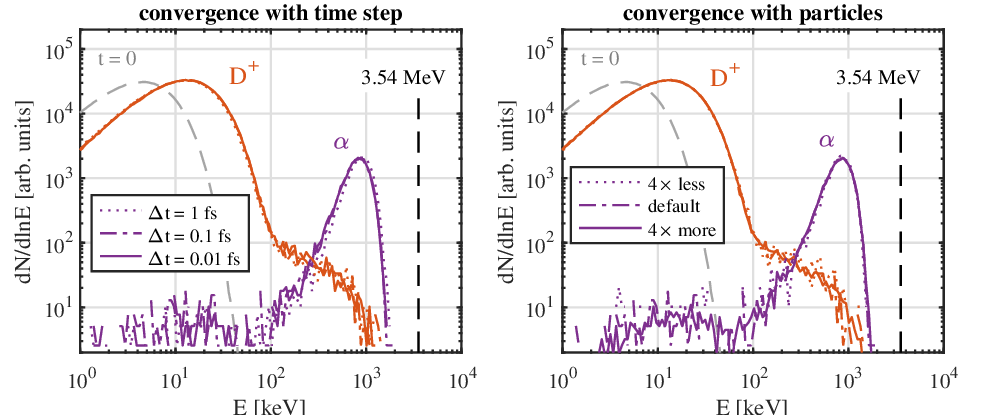}
\caption{Convergence of $D$ and $\alpha$ energy distributions at $t=2$\,ps with time step (left) and particles (right). The results on the left are obtained using the default number of particles per cell. The results on the right are obtained using $\Delta t = 0.1$\,fs.} \label{fig:alphaBeam_convergencTimeStepAndNppc}
\end{figure}

As with any numerical algorithm, it is important to show convergence with numerical parameters. Convergence with time step and particles is illustrated in Fig.\,\ref{fig:alphaBeam_convergencTimeStepAndNppc} by examining the energy distribution of the D and $\alpha$ species at $t=2$\,ps. The results are qualitatively correct using $\Delta t = 1$\,fs, but there are quantitative differences with $\Delta t = 0.1$\,fs. There is no discernible difference when decreasing the time step another factor of 10 to $\Delta t = 0.01$\,fs. The default particle resolution is also shown to be sufficient to quantify the knock-on production.

\section{Discussion and summary} \label{sec:summary}

A binary-pairing collision method is introduced for screened Coulomb collisions in plasmas that is valid in the weakly and moderately coupled regimes. While the binary-pairing procedure follows previous works \cite{Takizuka1977205}\cite{angus2024mpmccc}, the novelty lies in the approach for determining the polar scattering angle. The method introduced here essentially performs single Rutherford scattering for a range of impact parameters that is reasonably possible given the simulation time step. The cumulative distribution is used to maintain the correct transport cross section when the simulation time step does not permit using $b_{c}=b_{\max}$. The method presented here can be integrated into existing binary-pairing methods for cumulative Coulomb scattering with the addition of about 10 lines of code.

Leading-order $1/\ln\Lambda$ corrections to Fokker-Planck transport \cite{Li1993} enter via the second transport moment \cite{Bobylev2013123}. These effects can be captured even for time steps such that $b_c<b_{\max}$. It is interesting to comment that 1) Bloch's correction to the Bethe stopping power formula is computed using the second transport moment, and 2) that the classical and quantum calculation of this quantity give identical results \cite{lindhard1996}. Large-angle scattering is entirely omitted for collisions involving electrons in Ref.\,\cite{TURRELL2015144}, attributed to quantum mechanical effects. This also means that $1/\ln\Lambda$ corrections to transport are absent in that model for electron interactions. In this work, such collisions are not excluded when using the large-angle scattering model. We use a modified relationship between the scattering angle and the impact parameter that yields a transport cross section in close agreement with the quantum mechanical calculation \cite{lindhard1996}. However, the model presented here is quasi-classical at best, and a fully quantum-mechanical approach is needed for an accurate treatment of large-angle scattering for high-velocity particles where $\kappa<1$.

The generalized Coulomb method that accounts for large-angle scattering is used to examine collisional stopping of high-energy alpha particles in a D/T plasma using ignition relevant parameters. Including large-angle scattering can significantly enhance the tails of the D/T energy distributions, but the stopping power is essentially the same as that for models that only consider cumulative scattering. It should be emphasized that this is mainly a benchmarking exercise since only quasi-classical Coulomb scattering is included. Accurately describing high-energy D/T ions produced by knock-on collisions with energetic alpha particles requires accounting for nuclear interference, which can enhance large-angle scattering \cite{Ballabio1997}. Future work will incorporate this effect, and the model will be used to study a wider range of plasma parameters, including regimes where $1/\ln\Lambda$ corrections to Fokker-Planck transport are important.

\section*{Acknowledgments}
This work was performed under the auspices of the U.S. Department of Energy by Lawrence Livermore National Laboratory under Contract DE-AC52-07NA27344 and was supported by the LLNL-LDRD Program under Project No. 23-ERD-007.

\appendix
\section{Relativistic generalizations} \label{appendixA}

The equations presented in this paper consider the non-relativistic limit for simplicity. The generalization of the relevant quantities for relativistic kinematics is given here. We start with the following two definitions for the relative velocity of particles $i\in\alpha$ and $j\in\beta$:
\begin{eqnarray}
v_{ij} &\equiv & |\textbf{v}_i - \textbf{v}_j|, \label{Eq:vR} \\
v_{ij,\text{inv}} &\equiv & \frac{\sqrt{|\textbf{v}_i - \textbf{v}_j|^2 - |\textbf{v}_i\times\textbf{v}_j|^2/c^2}}{1-\textbf{v}_i\cdot\textbf{v}_j/c^2}. \label{Eq:vR_inv}
\end{eqnarray}
Here, $v_{ij}$ is the standard definition of the relative velocity of particles $i$ and $j$, which is not a relativistic invariant. The definition in Eq.\,\ref{Eq:vR_inv}, on the other hand, is a relativistic invariant. The superscript $*$ is used below to denote quantities computed in the center-of-momentum (COM) frame. The relativistic generalization of $b_{\bot}$ and $b_{\text{qm}}$ are
\begin{eqnarray}
b_{\bot} &=& \frac{|q_{\alpha}q_{\beta}|}{4\pi\epsilon_0\mu_{\alpha\beta}v^2_{ij}} \rightarrow \frac{|q_{\alpha}q_{\beta}|}{4\pi\epsilon_0\mu^*_{\alpha\beta}v^*_{ij}v^*_{ij,\text{inv}}}, \label{Eq:bperp_rel} \\
b_{\text{qm}} &=& \frac{\hbar}{2\mu_{\alpha\beta}v_{ij}} \rightarrow \frac{\hbar}{2\mu^*_{\alpha\beta}v^*_{ij}}, 
\end{eqnarray}
where $\mu^*_{\alpha\beta}=\gamma_1^*m_{\alpha}\gamma_2^*m_{\beta}/\left(\gamma_1^*m_{\alpha}+\gamma_2^*m_{\beta}\right)$ is the relativistic definition of the reduced mass with $\gamma$ the particle relativistic factors. The momentum of either particle in the COM frame can be written as $|\textbf{p}^*_i|=|\textbf{p}^*_j|=\mu^*_{\alpha\beta}v^*_{ij}$. The relativistic generalization of $b_{\bot}$in Eq.\,\ref{Eq:bperp_rel} is consistent with the relativistic Rutherford cross section \cite{Frankel1979}, which is defined in the COM frame and invariant by definition \cite{landau2013classical}. The dimensionless parameter $\kappa = b_{\bot}/b_{\text{qm}}$, which determines the relative importance of quantum effects, is invariant. Finally, the generalizations of $N_{ij}$ and $s_{ij}$ in Eqs.\,\ref{Eq:Nij}-\ref{Eq:sij}, which are also relativistic invariants, are
\begin{eqnarray}
N_{ij} &=& \sigma_{\text{tot}}v_{ij}n_{\text{min}}\Delta t \rightarrow \sigma_{\text{tot}}v^*_{ij}n_{\text{min}}\Delta t \frac{\gamma^*_1\gamma^*_2}{\gamma_1\gamma_2},\\
s_{ij} &=& \sigma_{\text{tr}}v_{ij}n_{\text{min}}\Delta t \rightarrow \sigma_{\text{tr}}v^*_{ij}n_{\text{min}}\Delta t \frac{\gamma^*_1\gamma^*_2}{\gamma_1\gamma_2}.
\end{eqnarray}
Here, the effective density $n_{\min}$ and time step $\Delta t$ are computed in the lab/simulation frame, same as $\gamma_1$ and $\gamma_2$.

\bibliographystyle{model1-num-names}

\end{document}